# Time-Domain Bound States in the Continuum

Oded Schiller[1,3], Yonatan Plotnik[3], Ohad Segal[1,3], Mark Lyubarov[2,3] and Mordechai Segev[1,2,3]

1. *Department of Electrical and Computer Engineering, Technion, Haifa 32000, Israel*
2. *Physics Department, Technion, Haifa 32000, Israel*
3. *Solid State Institute, Technion, Haifa 32000, Israel*

**Abstract**

We present the concept of time-domain bound states in continuum. We show that a rapid judiciously-designed temporal modulation of the refractive index in a spatially homogenous medium gives rise to a bound state in time embedded in a continuum of wavenumbers. Mathematically, these bound states in the continuum (BIC) are analytic solutions of the Maxwell equations in time and one-dimensional space. Our results show the potential to extend known wave phenomena in space to the temporal domain, providing new avenues for light-matter interactions in time-varying media.

Generally, a quantum system described by the Schrödinger equation with a finite potential well displays two kinds of eigenstates: a finite number of bound states and a continuum of unbound states, depending on whether the state's energies (eigenvalues of the linear Schrödinger equation) lie below or above the level of the ambient potential. A Bound State in the Continuum (BIC) is a special eigenstate of a system, which has energy above the continuum threshold, but counterintuitively, the state is bound. The idea of BICs was first proposed in 1929 [1] by J. von Neuman and E. Wigner. They found a special potential structure which supports a bound eigenstate, with energy that is higher than the ambient level of the potential, a bound state residing in the continuum of unbound modes. Over the years, the concept of BICs was further extended to a variety of other wave systems, such as electromagnetic (EM) waves [2] sound waves [3] and others. The idea was also shown theoretically in more quantum system, for example ones that have separable Hamiltonian [4] and for two electron systems [5]. Until recently, the idea of a BIC was only a theoretical concept, because the BICs found by von Neuman and Wigner has an infinite support in space. Namely, the potential is decaying in an oscillating fashion ad infinitum. Trimming this potential immediately couples the state to the continuum modes, and it is no longer a bound state. For this reason, early attempts to observe BICs in experiments focused on bound states above a potential well (but not in the continuum) [6] and alike. More recently, BICs were proposed to exist also in potentials other than the one proposed by von Neuman & Wigner, e.g., exploiting particular symmetries in the system. This made it possible to observe BICs experimentally, which was realized in 2011 [7], for paraxial EM waves, where a specific symmetry in an array of evanescently-coupled waveguides produced a BIC. One might add that a BIC like state was also shown in acoustic wave system beforehand [8], even though it was not known at the time of writing. Since then, many examples of BICs were observed in experiments in a wide range of wave systems [9,10], including BICs arising from accidental symmetries and BICs with topological properties [11]. The interest in BICs, especially after they were demonstrated in experiment, is partly fueled by their unique properties. One can think of a BIC as a resonant state. In general, a resonant state is a partially localized state with a finite lifetime, whose energy resides within the continuum [12]. In this sense, the BIC is a resonant state with infinite lifetime. This property means, for example, that a BIC can carry finite energy despite being surrounded by unbound states and can be maintained with zero loss that would otherwise be introduced by coupling to the radiation modes [13–15]. However, thus far, all BICs found theoretically or observed in experiments were bound states in space, never BICs in time.

Here, we present the first BICs bound in the time dimension. We solve the Maxwell equations for a spatially-homogenous medium where the refractive index is varied in time, and find a specific temporal profile of the refractive index that supports an EM wave localized in the time domain. For such time-domain BICs, the Poynting vector is absolute integrable (i.e., the time-integral of the absolute value of the Poynting vector is finite) and the refractive index approaches unity at infinite time. Interestingly, at the same singular point in the continuum spectrum we find an additional state that is exponentially rising in time: an "Anti-BIC". While this "Anti BIC" reflects the generic nature of $2^{nd}$ order PDEs, it is generally ignored since energy conservation renders it unphysical. In the case of time modulation, the energy restriction is lifted – because the Anti-BIC can extract energy from the modulation. These findings inspire further investigation of BICs in time-varying systems, and in space-time systems. A temporal BIC provides insight to the fundamental physics of time-varying media, and opens new avenues for light-matter interactions therein, topics that garner extensive interest in recent years.

Since the temporal BIC is supported by a time-varying permittivity, a short introduction on EM waves in time-varying media is due. This topics has been receiving growing attention in recent years, partly due to recent experimental advances in the ability to dramatically change the refractive index of some materials on the time scale of a single cycle, both at optical frequencies [16] and in microwaves [17,18]. One of the most prominent phenomena in this field is the photonic time-crystal (PTC) [19,20], a spatially homogenous medium where the refractive index changes periodically in time (or a spatio-temporal photonic crystal [21,22]). The PTC is the temporal analogue of the "conventional" photonic crystal (PC): a time invariant medium with a refractive index that varies periodically in space. Comparing the solutions of Maxwell's equations for a PC and a PTC, especially in the context of light-matter interactions, gives rise to a plethora of new ideas, ranging from light emission and generation of entangled pairs of photons to interactions with free electrons and more [23,24].

We begin by considering a spatially-homogeneous dielectric medium, where the refractive index changes in time. We are interested in an EM field that depends only on one spatial dimension, which we denote as z. We assume the field is linearly polarized in the x direction, as illustrated in Fig. 1. Under these assumptions, Maxwell's equations lead to the following equation for the electric displacement field, *D*:

$$(n(t)^2 \partial_t^2 - c^2 \partial_z^2)\vec{D}(t,z) = 0 \qquad (1)$$

Where $n(t)$ is the refractive index and $c$ is the speed of light in vacuum. For a non-magnetic dielectric medium, as we assume henceforth, $\epsilon(t) = n(t)^2$, where $\epsilon(t)$ is the relative permittivity. This assumption is not necessary but simplifies the derivation.

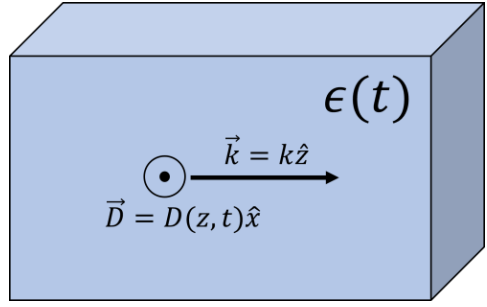

*Fig. 1. Geometry of the system. An x-polarized EM field with wavevector k pointing in the z direction is traveling in a uniform time-varying medium.*

Next, we assume that the field has a well-defined wave-vector, which we denote as $\vec{k} = k\hat{z}$. Because the medium is spatially-invariant, the wave-vector is conserved. This implies that the solution can be chosen to be of the form

$$\vec{D}(t,z) = e^{ikz} \cdot T(t)\hat{x} \qquad (2)$$

Where $T(t)$ is a function of time but does not depend on space. For a state to be a BIC in time, the electric and magnetic fields, as well as the Poynting vector, must be finite everywhere and at any time, without singularities. The Poynting vector must be integrable, so that the energy carried by the BIC would be finite [1]. These criteria assure that the state is bound. For the state to be in the continuum, we also require the refractive index be finite and real at any time, and to decay to unity at infinite time. This requirement ensures that there is a continuum of states, since far enough in time the refractive index is the same as in vacuum, up to an arbitrarily small variation, and of course in vacuum we have continuum of states [25].

After laying the ground for what a BIC in time means mathematically, we now explain how we find an analytical solution to this problem. To find a BIC in time, we follow in the footsteps of [1]: we assume a wave-function and find a potential (dielectric profile) that supports it. We use the following parametric family of analytic solutions

$$T(t) = \sin(\Omega t) \cdot \exp\left(-\gamma \int_0^{\Omega t} \frac{y^{-r} \cos^{\beta-1}(y) \sin^{\alpha+1}(y)}{1 + y^{-r} \cos^\beta(y) \sin^\alpha(y)} dy\right) \tag{3}$$

Where $\Omega = ck$, $\alpha, \beta, \gamma, r$ are parameters that for a wide range of values give a bound state. This ansatz is essentially a BIC solution to a slowly-varying envelope approximation of Eq. (1), which is mathematically equivalent to the Schrödinger equation (see supplementary information). For simplicity, we define the integrand as

$$f(y) = \frac{\gamma \cdot y^{-r} \cos^{\beta-1}(y) \sin^{\alpha+1}(y)}{1 + y^{-r} \cos^\beta(y) \sin^\alpha(y)} \tag{4}$$

Even though this wave-function ansatz is a result of an approximation, we are able to find a dielectric profile supporting it for the full wave equation, Eq. (1):

$$\epsilon(t) = \frac{1}{1 + 2\cot(\Omega t) f(\Omega t) + f'(\Omega t) - f(\Omega t)^2} \tag{5}$$

For the parametric family shown above, we find a BIC in time for a wide range of parameters selections. Henceforth we focus on two representative solutions that represent two interesting types of solutions for which the asymptotic behavior of the field and the Poynting vector can be calculated analytically.

The first solution is a BIC, under the following choice of parameters, $\gamma = \frac{1}{4}, \beta = 1, \alpha = 1, r = \frac{1}{2}$. This state has an absolute integrable Poynting vector (Fig. 2(c)), and the refractive index supporting it (Fig. 2(b)) decays to unity at infinite times. We find the exact asymptotic behavior of the envelope of the electric displacement field, $D$, by calculating the behavior analytically (for more details see Supplementary Information). As shown in Fig. 2(a), the field has an oscillatory component with a decaying envelope. The field oscillates at frequency $\Omega = ck$, similar to the vacuum dispersion relation. We find the asymptotic behavior for large times ($t \gg 1/\Omega$, the time scale of the state). The asymptotic behavior (Fig. 2(d)) for large time scales is $D \propto e^{-\frac{1}{4}\sqrt{\Omega t}}$, which is indeed square-integrable. In a similar vein, $\epsilon(t)$ supporting this state has decaying oscillations that converge to $\epsilon(t) = 1$, with the asymptotic behavior $\epsilon(t) - 1 \propto (\Omega t)^{-\frac{1}{2}}$ (Fig. 2(e)). The instantaneous Poynting vector is shown in Fig. 2(c), and its asymptotic behavior (Fig. 2(f)) is $S(t) \propto e^{-\frac{1}{2}\sqrt{\Omega t}}$, which is integrable. By integrating the energy flux over time, we find that the energy per unit volume is finite, making the state a time-BIC.

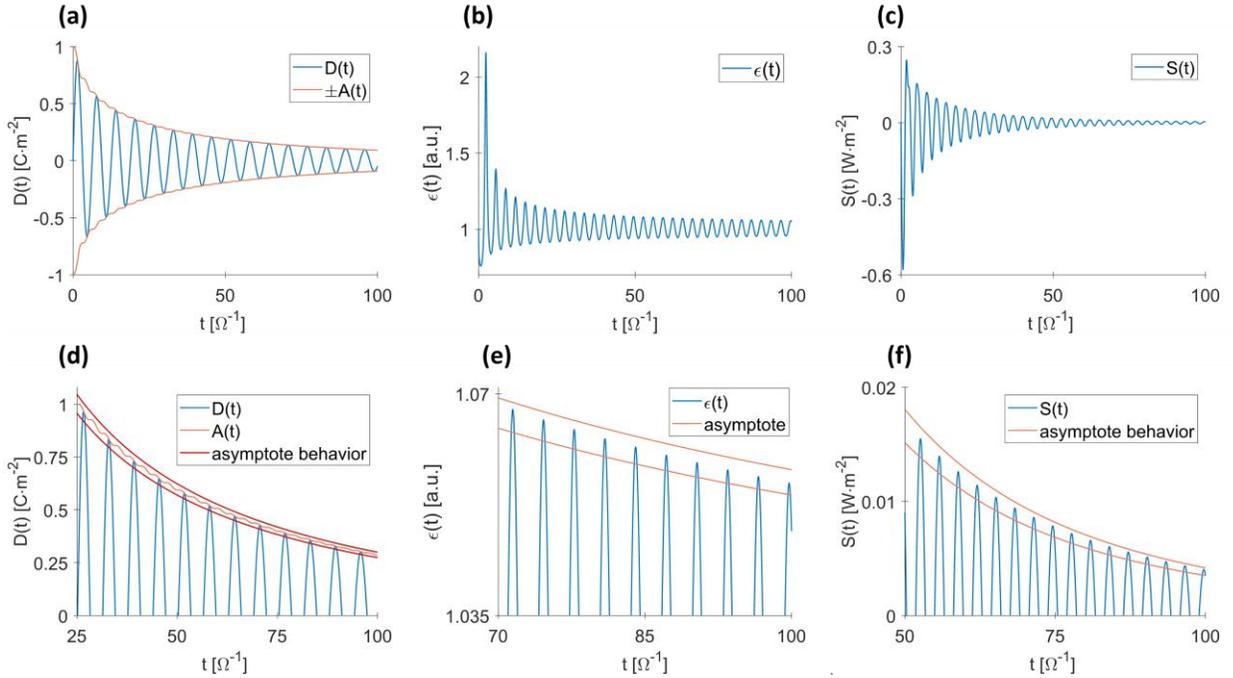

*Fig. 2. BIC solution with parameters $\gamma = \frac{1}{4}, \beta = 1, \alpha = 1, r = \frac{1}{2}$. (a) The electric displacement field D field of the state, and the field envelope (orange). (b) $\epsilon(t)$ generating this state. (c) Instantaneous Poynting vector. (d) Asymptotic behavior of D. (e) Asymptotic behavior of $\epsilon(t)$. (f) Asymptotic behavior of $S(t)$*

This BIC is localized in time, and is somewhat similar to the topological temporal edge state between two PTC of different Zak phases [26]. The topological edge state starts with zero electric field at t→-∞, increases its amplitude until t=0, and then decays back to zero electric field at t→∞ [19]. However, the topological edge state is not a BIC, because its wavenumber resides in the (topological) gap, and is not embedded in a continuum of modes, as is the BIC we present in Fig. 2.

The second solution (Fig. 3) occurs under the parameters $\gamma = \frac{1}{2}, \beta = 1, \alpha = 1, r = 1$. Apart from also being embedded in a continuum of modes, this state behaves differently than the BIC of Fig. 2. We calculate analytically the asymptotic behavior of this solution and find that the envelope of the electric displacement field for large times acts as $D \propto (\Omega t)^{-\frac{1}{4}}, \epsilon(t) - 1 \propto (\Omega t)^{-1}$ and the instantaneous Poynting vector behaves as $S(t) \propto (\Omega t)^{-\frac{1}{2}}$ (for details see Supplementary Information). This means that the Poynting vector is not integrable: it does decay, but too slowly to be integrable, hence it carries infinite energy per unit volume, and therefore this state is not a BIC.

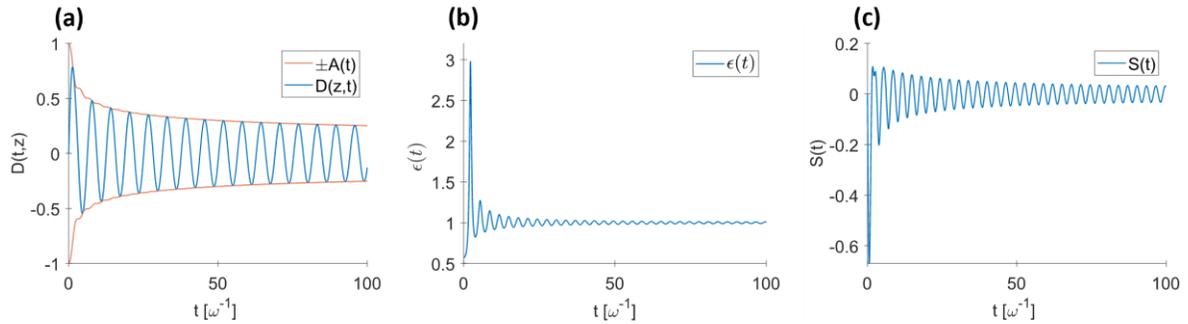

*Fig. 3. The state formed under parameters $\gamma = \frac{1}{2}, \beta = 1, \alpha = 1, r = 1$. (a) Electric displacement field D, and its envelope (orange). (b) $\epsilon(t)$ generating this state. (c) Instantaneous Poynting vector.*

The difference between the BIC (Fig. 2) and the solution of Fig. 3 is interesting. We can see that for the $\epsilon(t)$ that decays faster, $D$ and the Poynting vector decay slower, and the vice versa. This seems to be a general feature of this family of the solutions we find for $\epsilon(t)$ defined by Eq. 5, not only of the two examples we show. This is somewhat analogous to quantum mechanics, where generally - if the potential decays faster - the wave function is more spread out. In our case, $\epsilon(t)$ plays the role of the potential and **D** is the eigenfunction.

An immediate consequence of the existence of a decaying mode (such as the two solutions presented in Figs 2, 3) is the presence of its twin: a rising mode with amplitude increasing in time. This mode has exactly the same wavevector but its phase is shifted by $\pi/2$. When considering a spatial BIC, this mode is ignored, since it carries infinite energy and therefore not physical. However, in time-varying systems such as ours, energy is not conserved (because time-translation is broken by the modulation), hence the raising state can exist, drawing energy from the modulation of $\epsilon(t)$. We call such a state an "Anti-BIC. Figure 4 shows **D(t)** of the Anti-BIC for the parameters of Fig.2, $\gamma = \frac{1}{4}, \beta = 1, \alpha = 1, r = \frac{1}{2}$. For this Anti-BIC, we find a singular point in the continuum spectrum which displays an exponential rise, in contradistinction with the topological edge state between two PTCS, which resides in a finite gap in momentum that can support exponential gain [26] for a continuous segment of $k$ values.

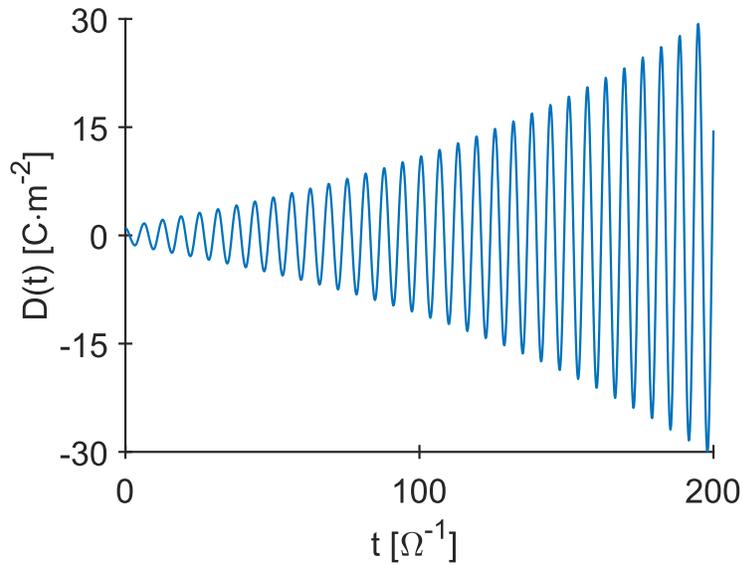

*Fig. 4. Electric displacement field D of the "Anti-BIC" under parameters $\gamma = \frac{1}{2}, \beta = 1, \alpha = 1, r = 1.$*

Thus far, we discussed the time interval from t=0 to t→∞, but for the BIC to be truly bound, $\boldsymbol{D}$ and $\boldsymbol{S}$ must also decay as we approach negative infinity, to make it a state that is localized in time. To this end, we know that Maxwell's equations are time-reversible, so if we take $\epsilon(t)$ to be symmetric in time, and $\boldsymbol{D}$ to be anti-symmetric in time, we get a solution to Maxwell's equations that is valid at all times. In this vein, we find a mode that rises from $t \to -\infty$, until $t = 0$, and then decays towards $t \to \infty$. This yields a mode that is localized in time, and a true BIC. Of course, the asymptotic behavior for $t \to \infty$ still holds for $t \to -\infty$, and we get an absolute integrable Poynting vector. Notice that the choice of $D = 0$ at $t = 0$ ensures that no energy is coupled into the Anti-BIC.

To summarize, we presented the concept of time-domain bound states in the continuum, found an analytic family of parametric solutions that for a large variety of parameters yield a BIC in time. We studied two specific solutions highlighting two different behaviors. With the recent experimental progress in observing time-reflections in microwaves [17,18] and in observing time-refraction within a the single-cycle at optical frequencies [16] we hope that an experimental demonstration of a time-domain BIC will follow shortly, since it can be achieved in a variety of wave systems.

Last but not least, we note one big theoretical question raised by this work: what happens at the quantum level when a time-domain BIC is excited. Answering this question in the quantum realm can shed new light on the basic physics of time varying medium, and inspire new kinds of light-matter interactions in the presence of a time-domain BIC playing the role of "confining potential".

Another open question is the existence of a spatio-temporal BIC: a BIC bound in both space and time.

This work was supported by the Breakthrough Program of the Israel Science Foundation and by a grant from the US Air Force for Scientific Research (AFOSR).

**References:**

[1] J. von Neuman and E. Wigner, *Uber Merkwürdige Diskrete Eigenwerte. Uber Das Verhalten von Eigenwerten Bei Adiabatischen Prozessen*, Physikalische Zeitschrift **30**, 467 (1929).

[2] D. C. Marinica, A. G. Borisov, and S. V. Shabanov, *Bound States in the Continuum in Photonics*, Phys. Rev. Lett. **100**, 183902 (2008).

[3] C. M. Linton and P. McIver, *Embedded Trapped Modes in Water Waves and Acoustics*, Wave Motion **45**, 16 (2007).

[4] M. Robnik, *A Simple Separable Hamiltonian Having Bound States in the Continuum*, J. Phys. A: Math. Gen. **19**, 3845 (1986).

[5] F. H. Stillinger and D. R. Herrick, *Bound States in the Continuum*, Phys. Rev. A **11**, 446 (1975).

[6] F. Capasso, C. Sirtori, J. Faist, D. L. Sivco, S.-N. G. Chu, and A. Y. Cho, *Observation of an Electronic Bound State above a Potential Well*, Nature **358**, 6387 (1992).

[7] Y. Plotnik, O. Peleg, F. Dreisow, M. Heinrich, S. Nolte, A. Szameit, and M. Segev, *Experimental Observation of Optical Bound States in the Continuum*, Phys. Rev. Lett. **107**, 183901 (2011).

[8] R. Parker, *Resonance Effects in Wake Shedding from Parallel Plates: Some Experimental Observations*, Journal of Sound and Vibration **4**, 62 (1966).

[9] C. W. Hsu, B. Zhen, J. Lee, S.-L. Chua, S. G. Johnson, J. D. Joannopoulos, and M. Soljačić, *Observation of Trapped Light within the Radiation Continuum*, Nature **499**, 188 (2013).

[10] C. W. Hsu, B. Zhen, A. D. Stone, J. D. Joannopoulos, and M. Soljačić, *Bound States in the Continuum*, Nat Rev Mater **1**, 9 (2016).

[11] B. Zhen, C. W. Hsu, L. Lu, A. D. Stone, and M. Soljačić, *Topological Nature of Optical Bound States in the Continuum*, Phys. Rev. Lett. **113**, 257401 (2014).

[12] L. D. Landau and L. M. Lifshitz, *Quantum Mechanics: Non-Relativistic Theory*, 3rd edition (Butterworth-Heinemann, Singapore, 1981).

[13] H. Friedrich and D. Wintgen, *Interfering Resonances and Bound States in the Continuum*, Phys. Rev. A **32**, 3231 (1985).

[14] L. Fonda and R. G. Newton, *Theory of Resonance Reactions*, Annals of Physics **10**, 490 (1960).

[15] S. Fan, W. Suh, and J. D. Joannopoulos, *Temporal Coupled-Mode Theory for the Fano Resonance in Optical Resonators*, J. Opt. Soc. Am. A, JOSAA **20**, 569 (2003).

[16] E. Lustig et al., *Time-Refraction Optics with Single Cycle Modulation*, Nanophotonics **12**, 2221 (2023).

[17] T. R. Jones, A. V. Kildishev, M. Segev, and D. Peroulis, *Time-Reflection of Microwaves by a Fast Optically-Controlled Time-Boundary*, arXiv:2310.02377.

[18] H. Moussa, G. Xu, S. Yin, E. Galiffi, Y. Ra'di, and A. Alù, *Observation of Temporal Reflection and Broadband Frequency Translation at Photonic Time Interfaces*, Nat. Phys. **19**, 6 (2023).

[19] J. R. Reyes-Ayona and P. Halevi, *Observation of Genuine Wave Vector (k or β) Gap in a Dynamic Transmission Line and Temporal Photonic Crystals*, Applied Physics Letters **107**, 074101 (2015).

[20] F. Biancalana, A. Amann, A. V. Uskov, and E. P. O'Reilly, *Dynamics of Light Propagation in Spatiotemporal Dielectric Structures*, Phys. Rev. E **75**, 046607 (2007).

# Supplementary Information for "Time-Domain Bound States in the Continuum"

### A. Reasoning for the ansatz in Eq. 3 of the main text.

Equation (3) presented an ansatz that we use to find the BIC that solves Eq. 1, which is a wave equation for D derived from Maxwell's equations in a spatially homogeneous medium that can vary in time. In this section, we explain how we come up with this ansatz. If we substitute $\vec{D}(t, z)$ of the form

$$\vec{D}(t, z) = e^{ikz} T(t) \hat{x}$$

into Eq.1, where we use a state with a well-defined wavenumber, k, we get the following equation for $T(t)$, i.e. the time-dependent part of $D$:

$$\epsilon(t) T''(t) + \Omega^2 T(t) = 0 \qquad . \qquad (S1)$$

This equation can be handled in two ways. The first is the conventional way those kinds of equations are solved. Namely, for a given dielectric function $\epsilon(t)$, we can find which eigenstates, $T(t)$, it supports. The second way we can handle such an equation is in the spirit it was handled in Ref. [1]. Namely, for a given state, $T(t)$, we can construct a dielectric function $\epsilon(t)$ that can support it. We shall use the second method. This is because a BIC arises only for a very specific $\epsilon(t)$ and specific initial conditions, so taking the first route we would not only need to find the specific $\epsilon(t)$ that can support a BIC, but we would also need to find the specific initial conditions that define it. It is like finding a needle in a haystack. On the other hand, when using the second method, stating with a state with the desired attributes and finding a dielectric function $\epsilon(t)$ that supports it, there is no problem with initial conditions, because only one specific $\epsilon(t)$ will support each state.

Having explained why we are trying to find a (real) function $\epsilon(t)$ from a guess of the electric field, we will now explain how we actually find the initial guess given in Eq. (3).

We start by using the following form for $T(t)$

$$T(t) = A(t)\sin(\Omega t) \tag{S2}$$

which consists of an oscillatory function $\sin(\Omega t)$ with the frequency matching the vacuum dispersion relation ($\Omega = ck$), and a time-dependent envelope $A(t)$. We use this form of solution because we expect that at time scales shorter than the decay time of the BIC, the solution will approximately coincide with a plane wave, which is the solution of Maxwell's equations in a homogeneous medium with a specific wavenumber. We are facing the challenge of finding a waveform that decays fast enough in time (such that it has absolute integrable Poynting vector), for which we can analytically find a dielectric function $\epsilon(t)$ that is real, finite, and positive at all times, and decays to a constant value at infinite time. We find that, for most waveform choices, these requirements cannot be fulfilled.

To find the BIC solution for the full equation (S1), we first obtain an initial guess for the approximated version of Eq. S1, under the slowly-varying envelope approximation Under this approximation, valid when the second derivative of $A(t)$ is very small, we find:

$$A'_{ap}(t) - \frac{\Omega}{2}\frac{\epsilon_{ap}(t) - 1}{\epsilon_{ap}(t)}\tan(\Omega t)\, A_{ap}(t) = 0 \quad . \tag{S3}$$

Here, $A_{ap}(t)$ is the envelope function supported by $\epsilon_{ap}(t)$, under the slowly-varying approximation for $A(t)$. This a first order ODE, so we can find a closed-form solution for $A_{ap}(t)$, for each $\epsilon_{ap}(t)$ we use. We use the notation $A_{ap}(t)$ and $\epsilon_{ap}(t)$ to emphasize that those are not yet the final $A(t)$, $\epsilon(t)$ we will find, because we want to find a solution for Eq. 1, which is the full (classical) wave equation. To do that, we use a slightly modified version of $A_{ap}(t)$, and find a dielectric function $\epsilon(t)$ and that solves Eq S1. Implementing the procedure stated above, we first solve Eq. S3. For that, we choose

$$\epsilon_{ap}(t) = 1 - (\Omega t)^{-r} \cos^\beta(\Omega t) \sin^\alpha(\Omega t)$$

which is a PTC that decays in time polynomially. The intuition behind this choice is that the modes of PTC momentum gap rise/decay exponentially, with a rate proportional to the amplitude of the refractive index modulation. This means that, as the amplitude diminishes - so does the decay rate of the modes and the gap shrinks. Thus, if we choose the decay rate properly (slowly enough), we can find a single mode in the middle of the gap that decays fast enough to be square-integrable: a BIC. By substituting $\epsilon_{ap}(t)$ into Eq. S3, we find the following solutions for $A_{ap}(t)$

$$A_{ap}(t) = \exp\left(-\frac{1}{2}\int_0^{\Omega t} \frac{y^{-r} \cos^{\beta-1}(y) \sin^{\alpha+1}(y)}{1 - y^{-r} \cos^\beta(y) \sin^\alpha(y)} dy\right) \quad . \tag{S4}$$

For many choices of parameters, this approximation has the desired attributes of a BIC. The initial guess $\epsilon_{ap}(t)$ also has the proper form to get a BIC, as it decays to unity when $t \to \infty$ and is real finite and positive. The hope is that, when we find the $\epsilon(t)$ that supports this initial guess, for the full equation, it will be similar to $\epsilon_{ap}(t)$, and preserve the desired attributes it has. If this assumption holds, then we have a BIC solution for the full wave equation (Eq. S1).

We now find $\epsilon(t)$ that can support the state we have found under the approximation, in the full (S1) equation. We use a slightly modified version of $A_{ap}(t)$:

$$T(t) = \sin(\Omega t) \cdot \exp\left(-\gamma \int_0^{\Omega t} \frac{y^{-r} \cos^{\beta-1}(y) \sin^{\alpha+1}(y)}{1 - \delta y^{-r} \cos^\beta(y) \sin^\alpha(y)} dy\right) \tag{S5}$$

Which is the ansatz we showed in Eq (3) of the main text. Notice that we added two parameters, $\gamma, \delta$. These additional parameters are added to make sure that the $\epsilon(t)$ of the original wave equations, satisfies all the wanted attributes. The $\delta$ parameter is mostly added to make sure that $\epsilon(t)$ does not go to zero at t=0, which might happen for some parameter choices, in practice choosing $\delta = -1$ solves this issue, hence we use $\delta = -1$ throughout the paper. Substituting this ansatz in the full (S1) equation, we get the following solution for $\epsilon(t)$ that can support this state in Eq. S1. For simplicity, we define the integrand as

$$f(y) = \frac{\gamma \cdot y^{-r} \cos^{\beta-1}(y) \sin^{\alpha+1}(y)}{1 + y^{-r} \cos^{\beta}(y) \sin^{\alpha}(y)} \quad (S6)$$

The $\epsilon(t)$ of Eq. 1 is therefore

$$\epsilon(t) = \frac{1}{1 + 2\cot(\Omega t) f(\Omega t) + f'(\Omega t) - f(\Omega t)^2} \quad (S7)$$

as shown in the main text in equations (4) and (5).

We now have a parametric family of solutions for the full equation (Eq. 1), that for a wide range of parameters choices gives a BIC solution. Meaning that the field is decaying, the Poynting vector is integrable and $\epsilon(t)$ is real, positive, finite at any time, and decays to unity when $t \to \infty$.

### B. Asymptotic behavior

We now show the calculation for the asymptotic behavior of the two representative solutions described in depth in the main text.

Let us first look at the envelope of the electric displacement field, $D$, of the parametric family of solutions we found, $A(t) = \exp\left(-\gamma \int_0^{\Omega t} \frac{y^{-r} \cos^{\beta-1}(y) \sin^{\alpha+1}(y)}{1+y^{-r} \cos^{\beta}(y) \sin^{\alpha}(y)} dy\right)$. Let us calculate the asymptotic behavior of the integrand, $f(y)$, when $y \gg 1$. Under these conditions, we get, in the denominator, that $1 \gg y^{-r} \cos^{\beta}(y) \sin^{\alpha}(y)$, so we can say that for large times:

$$\gamma \int_0^{\Omega t} \frac{y^{-r} \cos^{\beta-1}(y) \sin^{\alpha+1}(y)}{1 + y^{-r} \cos^\beta(y) \sin^\alpha(y)} dy \approx \gamma \int_0^{\Omega t} y^{-r} \cos^{\beta-1}(y) \sin^{\alpha+1}(y) \, dy \qquad (S8)$$

Now, to calculate the asymptotic behavior for the state under parameters $\gamma = \frac{1}{4}, \beta = 1, \alpha = 1, r = \frac{1}{2}$ (the true BIC state), the approximate integral becomes $\frac{1}{4} \int_0^{\Omega t} y^{-\frac{1}{2}} \sin^2(y) \, dy$.

This integral has a closed-form solution

$$\frac{1}{4} \int_0^{\Omega t} y^{-\frac{1}{2}} \sin^2(y) \, dy = \frac{1}{4} \left( \sqrt{\Omega t} - \frac{\sqrt{\pi}}{2} C\left(\frac{2\sqrt{\Omega t}}{\sqrt{\pi}}\right) \right) \qquad (S9)$$

Where C(x) is the Fresnel cosine integral, which has a limit at $t \to \infty$. Thus, the asymptotic behavior of the envelope of $D$ is

$$A(t) = \exp\left( -\frac{1}{4} \int_0^{\Omega t} \frac{y^{-1/2} \sin^2(y)}{1 + y^{-1/2} \cos(y) \sin(y)} dy \right) \propto \exp\left( -\frac{1}{4} \sqrt{\Omega t} \right) \qquad (S10)$$

which is square-integrable.

For the other state, the one with the choice of parameters $\gamma = \frac{1}{2}, \beta = 1, \alpha = 1, r = 1$, we find that it is decaying but has infinite energy per unit volume, meaning that it is not a BIC. For that state, we get the integral $\frac{1}{2} \int_0^{\Omega t} y^{-1} \sin^2(y) \, dy$, which also has a closed-form solution

$$\frac{1}{2} \int_0^{\Omega t} y^{-1} \sin^2(y) \, dy = \frac{1}{4} (\ln(2\Omega t) - Ci(2\Omega t) + \Gamma) \qquad (S11)$$

Where $Ci(x)$ is the Cosine integral, which has a limit at $t \to \infty$, and $\Gamma$ is the Euler-Mascheroni constant. Thus, the asymptotic behavior of the envelope of $D$ is

$$A(t) = \exp\left( -\frac{1}{2} \int_0^{\Omega t} \frac{y^{-1} \sin^2(y)}{1 + y^{-1} \cos(y) \sin(y)} dy \right) \propto (\Omega t)^{-1/4} \qquad (S12)$$

which is not square-integrable.

Next, we examine the asymptotic behavior of the Poynting vector. We do that by finding the electric and magnetic fields via Maxwell's equations, and showing that the envelope of the Poynting vector scales as $A(t)^2$. In doing this, we find that, for the BIC state of parameters $\gamma = \frac{1}{2}$, $\beta = 1, \alpha = 1, r = 1$ the Poynting vector scales as $S \propto \exp\left(-\frac{1}{2}\sqrt{\Omega t}\right)$, which is integrable, as required from a BIC. But, for the state that decays too slowly to be square-integrable and is therefore not a BIC (parameters $\gamma = \frac{1}{2}, \beta = 1, \alpha = 1, r = 1$), we find that the Poynting vector scales as $S \propto (\Omega t)^{-1/2}$. The Poynting vector for this state is indeed not integrable, which means that the state has infinite energy per unit volume.

Finally, we find the asymptotic behavior of the dielectric function, for which we have found (Eq. 5 in the main text)

$$\epsilon(t) - 1 = -\frac{2\cot(\Omega t) f(\Omega t) + f'(\Omega t) - f(\Omega t)^2}{1 + 2\cot(\Omega t) f(\Omega t) + f'(\Omega t) - f(\Omega t)^2} \quad (S13)$$

We will calculate the asymptotic behavior for both states. When $\Omega t \gg 1$ the function $f(\Omega t)$ scales as $(\Omega t)^{-r}$. That means that $f(\Omega t)^2$ decays much faster: it decays as $(\Omega t)^{-2r}$, which means that it is negligible in comparison with $f(\Omega t)$ for $\Omega t \gg 1$. Because we have a closed-form expression for $f(y)$, we can find the derivative of this function, and get:

$$f'(y) = \frac{\sin(y)\,(2y^{r+1}\cos(y) + (y - ry^r)\sin(y))}{y(y^r + \sin(y)\cos(y))^2} \quad (S14)$$

The numerator scales up as $y^{r+1}$ and the denominator scales up as $y^{2r+1}$, which means that $f'(y)$ scales up as $y^{-r}$.

Using those calculations, we find that for these two states, $\epsilon(t) - 1$ scales as $(\Omega t)^{-r}$.